\documentclass[prl,twocolumn,showpacs]{revtex4}
\usepackage{amsfonts}
\usepackage{amsmath}
\usepackage{amssymb}
\usepackage{graphicx}
\usepackage{graphics}
\begin{document}

\noindent To be published in: \\ 
Phys. Rev. Lett. \textbf{93}, 039703 (2005) \\  

\noindent\textbf{Comment on ''Magnetic phase transition in 
Co/Cu/Ni/Cu(100) and Co/Fe/Ni/Cu(100)''} \\ 

In a recent Letter \cite{WWS03} the phase diagram of Co/Cu/Ni/Cu(100) 
and Co/Fe/Ni/Cu(100) trilayers was determined experimentally as well 
as theoretically. Whereas we do not object the presented results, 
we remark that published work was not commented on.
Especially, much more information can be 
extracted out of the measurements by a quantitative comparison with an 
appropriate model, allowing for a determination of important
parameters such as exchange coupling constants. In particular: 

\textbf{1.} For the theoretical investigation in \cite{WWS03} an 
Ising-type model has been assumed. To obtain a satisfactory agreement 
with the measured phase diagram a very large interlayer coupling 
$J_\mathrm{int}/J_\mathrm{Ni,Co}\sim0.1\ldots1$ as compared to the
ones of Ni and Co has been assumed, in disagreement with previous 
results \cite{BBK96,JBP99,WGK02}. However, we like to point out that 
this large value is caused by the use of Ising-like magnetic moments. 
By consideration of three-component Heisenberg spins, as appropriate 
for $3d$- transition-metal ferromagnets, much smaller values of 
$J_\mathrm{int}$ are sufficient to explain the measurements. The
reason for this large difference is the existence of collective 
magnetic excitations (spin waves), which are (i) absent for Ising-like 
spins, and which are (ii) particularly important for ultrathin films. 
%The latter is 
%caused by the close neighborhood of the weakly anisotropic 
%transition-metal ferromagnets to the Mermin-Wagner limit. 
Therefore, 
the resulting boundary values of $J_\mathrm{int}/J_{1,2}$ calculated 
in \cite{WWS03} are unphysically large. In addition, the shape of the 
magnetization $M(T)$ as a function of the temperature depends also 
sensitively on the type of spins involved. For Ising-like spins $M(T)$ 
is almost constant over a large temperature range and drops rapidly
to zero close to the Curie temperature. On the other hand, the
measured magnetization curve $M(T)$ can be described much better by 
three-component spins. 

Hence, for the description of the trilayer system a Heisenberg model 
should be applied, solved with improved approximations like a
many-body Green's function theory (GFT) \cite{FJK00} or Monte Carlo 
simulations \cite{SGL93}. As an example, in Fig.1 we present the Ni 
and Co magnetization curves as measured by XMCD \cite{SSB04} and 
calculated from GFT \cite{FJK00,JBP99}. In contrast, a simple mean 
field approximation neglects collective excitations and yields similar 
unphysical values for $J_\mathrm{int}$ as obtained by an Ising model. 

\textbf{2.} The boundaries between phases I-IV as presented in 
\cite{WWS03} are not always phase transition lines in the
thermodynamic sense. The ones between II-IV and III-IV are merely 
crossovers between thickness regions where magnetic domains in one of 
the ferromagnetic films can or cannot be detected. However, they are 
not accompanied by critical phenomena and by the (dis)appearance of 
an order parameter. As has been measured in \cite{BWS98} and
calculated in \cite{WWS03,WaM92,JBB00}, the corresponding 
susceptibility exhibits a maximum (``resonance'') at those boundaries 
but not a singularity. The phase transition line I-IV results if
this resonance and the ``real'' singularity merge into a single peak. 
Note furthermore that in addition to the crossover II-IV 
\cite{BWS98}, also the crossover III-IV \cite{SWB03} and the 
transition I-IV \cite{SWP01} have been reported previously. 

\begin{figure}[h] 
\includegraphics[width=6.5cm]{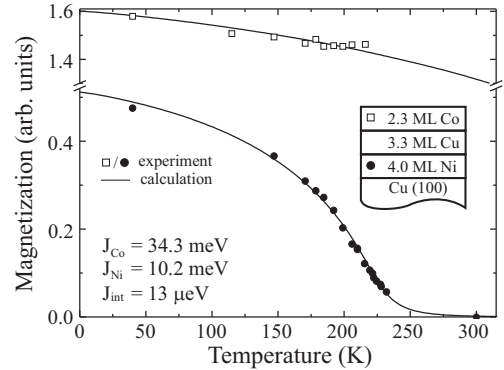} 
\caption{Ni and Co magnetizations of a Co/Cu/Ni/Cu(100) trilayer
  system as function of the temperature $T$. The measurements have
  been obtained by XMCD \cite{SSB04}. For the calculation we have
  assumed integer thicknesses next to the real ones, and exchange
  interactions as indicated.} 
\end{figure}

To conclude, we note that, in addition to the varying thicknesses of
the ferromagnetic layers, many more features of the trilayer system 
occur if also the thicknesses of the spacer and the cap layers are 
varied. Quite interestingly, the exchange couplings of the
ferromagnetic layers may depend sensitively on these thicknesses
\cite{Bru95}. A nonmonotonous behavior of the ordering temperature
may result \cite{BBK96}. \\ 
%Moreover, the interlayer exchange coupling 
%oscillates by varying the spacer thickness \cite{BWS98,PKT00}. \\

\noindent P. J. Jensen, C. Sorg, A. Scherz, M. Bernien, \\ K. Baberschke, 
and H. Wende \\
Physics Department, Freie Universit\"at Berlin, \\ 
Arnimallee 14, D-14195 Berlin, Germany \\[0.3cm] 
Received 27 February 2004 

\end{document}